\documentclass[12pt]{article}
\pdfoutput=1
\usepackage{graphicx,psfrag,epsf,color}
\usepackage{amsmath,amssymb,amsfonts,slashed}
\usepackage{array}
\usepackage{cite}
\usepackage{rotating}
\usepackage{slashed, cancel}
\usepackage{xcolor}
\usepackage[caption=false]{subfig}
\usepackage{graphicx}
\usepackage{mathrsfs}  
\usepackage{tabularx}
\usepackage{multirow} \usepackage{array}
\usepackage{floatrow}
\usepackage{physics}

\usepackage{color,xcolor}
\usepackage{lscape}
\usepackage[colorlinks,citecolor=blue,urlcolor=blue,linkcolor=blue]{hyperref}

\newcolumntype{C}{>{\centering\arraybackslash}X}
\numberwithin{equation}{section}
\begin{document}
\allowdisplaybreaks

\begin{titlepage}

\begin{flushright}
{\small
SI-HEP-2025-026\\
P3H-25-022\\
May 5, 2025 
}
\end{flushright}

\vskip1cm
\begin{center}
\Large \bf\boldmath Radiative tail of the \\  three-particle  light-cone distribution amplitudes \\ for the $\Lambda_b$ baryon in HQET
\end{center}

\vspace{0.5cm}
\begin{center}
Thorsten Feldmann, Daniel Vladimirov\\[6mm]

{\it Theoretische Physik 1, Center for Particle Physics Siegen, \\
Universit\"at Siegen, 57068 Siegen, Germany}\\[0.3cm]

\end{center}

\vspace{0.6cm}
\begin{abstract}
\vskip0.2cm\noindent

We calculate the so-called radiative tail for the three-particle light-cone distribution amplitudes (LCDAs) of the $\Lambda_b$ baryon, following from the short-distance expansion of the defining light-ray operators
in heavy-quark effective theory (HQET). 
To illustrate the effect of the radiative tail, we introduce a simple (model-dependent) extrapolation to large distances (respectively small momenta),
which preserves the analytic properties of the LCDAs in HQET.
We observe a similar qualitative behaviour as has been discussed for the $B$-meson case.

\end{abstract}

\end{titlepage}

\section{Introduction}
\label{sec:Intro}

Light-cone distribution amplitudes (LCDAs) for heavy $b$-hadrons appear in the context of the QCD factorization approach (QCDF \cite{Beneke:1999br,Beneke:2001ev}) and light-cone sum rules (LCSRs, see \cite{Khodjamirian:2023wol} and references therein) for exclusive energetic $B$-meson or $\Lambda_b$-baryon decays. 
The leading 2-particle LCDA for the $B$-meson has been studied extensively in the past. In particular, a measurement of the radiative semi-leptonic decay $B \to \gamma \ell \bar \nu_\ell$ 
at large photon energies can be used to constrain the $B$-meson LCDA,
see e.g.\ 
\cite{Korchemsky:1999qb,Descotes-Genon:2002crx,Lunghi:2002ju,Bosch:2003fc,Beneke:2011nf,Braun:2012kp,Wang:2016qii,Wang:2018wfj}.
Because of baryon-number conservation, an analogous decay mode that would allow for direct constraints on the $\Lambda_b$ LCDAs does not exist. As a consequence, so far, applications of QCDF \cite{Wang:2011uv} or LCSRs \cite{Feldmann:2011xf,Wang:2015ndk,Feldmann:2023plv} to heavy-baryon decays typically have to rely on certain model-dependent parameterizations that aim to give a reasonable description of the
momentum distribution of the light quarks inside the $\Lambda_b$ baryon.
For that reason, any kind of model-\emph{independent} information on the baryon LCDAs would be welcome. In fact, in heavy-quark effective theory (HQET) the distribution of \emph{large} light-cone momenta
is induced by short-distance dynamics and can be calculated in perturbation theory, leading to a power-like fall-off at large momenta \cite{Lange:2003ff,Lee:2005gza}. This so-called ''radiative tail'' arises because the heavy-quark in HQET has infinite mass and can thus transfer an \emph {infinite} amount of momentum to its light constituents.  
Formally, the radiative tail can be constructed from a short-distance expansion of 
the light-ray operators that appear in the field-theoretical definition of the LCDAs
\cite{Ball:2008fw,Ali:2012pn,Bell:2013tfa}. This has already been worked out for the $B$-meson case in \cite{Kawamura:2008vq}.

In this paper, we will present the results of the analogous calculation 
for the 3-particle LCDAs of the $\Lambda_b$ baryon in HQET, including operators 
of canonical mass dimesion $9/2$ and $11/2$ in the short-distance expansion of 3-quark operators in terms of two light-like distances $\tau_1$ and $\tau_2$ between the heavy and light quarks.
To this end, we will first introduce our setup and notation in Sec.~\ref{sec:setup}.
We will define the local operators and matching coefficients appearing  in the short-distance expansion of generic 3-particle baryonic light-ray operators in HQET in Sec.~\ref{sec:ope}.
In Sec.~\ref{sec:fixed-order} we will present our result for the 1-loop computation
of the matching coefficients, where we have followed the strategy that has been introduced in \cite{Feldmann:2023aml} for the case of $B_s$-meson LCDAs.
Evaluating the hadronic matrix elements of the local operators of dimension-9/2 and 11/2, we then derive the short-distance behaviour of the LCDAs in position space,  which can be expressed in terms of two ''decay constants'' and the HQET mass parameter $\bar \Lambda =M_{\Lambda_b}-m_b$. This is worked out in Sec.~\ref{sec:lcdas}, where we also propose a simple model-dependent extrapolation to large distances which -- after Fourier transformation to momentum space -- provides 1-loop improved results 
for the $\Lambda_b$-baryon LCDAs which include the perturbative constraints on the radiative tail to the considered order.
We conclude with a brief summary in Sec.~\ref{sec:conclusion}. 
Finally, for convenience, we summarize in an appendix the renormalization-group equations for the leading LCDA in momentum and position space.

\section{Setup and notation}
\label{sec:setup}

We introduce light-like vectors $n^\mu$ and $\bar n^\mu$, satisfying
$$
 n^2=\bar n^2=0 \,, \quad \mbox{and} \quad n\cdot\bar n= 2 \,,
$$
and choose a frame where the heavy-quark velocity is given by
$$
 v^\mu = \frac{n^\mu + \bar n^\mu}{2} \,.
$$
The heavy-quark field $h_v(x)$ in HQET fulfills the usual relations,
$$
 \slashed v \, h_v = h_v \,, \qquad (i v \cdot D) \, h_v =0 \qquad \mbox{(on-shell).}
$$
In the following, we will always consider the strict isospin limit for the light up- and down-quarks, ignoring QED corrections and setting $m_u=m_d \to 0$.

For the classification of the 3-particle LCDAs of the $\Lambda_b$ baryon in HQET, 
we follow  Ref.~\cite{Bell:2013tfa} (see also \cite{Ball:2008fw,Ali:2012pn}), and define 
\begin{align}
  \epsilon^{abc} \, \langle 0| \left(  u^a(\tau_1 n) \, 
  C \gamma_5 \slashed n  \,  d^b(\tau_2 n) \right) h_v^c(0) |\Lambda_b(v,s) \rangle 
  &= f_{\Lambda_b}^{(2)} \, \tilde \phi_2(\tau_1,\tau_2) \, u_{\Lambda_b}(v,s) \,, 
  \nonumber \\[0.2em]
   \epsilon^{abc} \, \langle 0| \left(  u^a(\tau_1 n) \, 
  C \gamma_5 \slashed {\bar n}  \,  d^b(\tau_2 n) \right) h_v^c(0) |\Lambda_b(v,s) \rangle 
  &= f_{\Lambda_b}^{(2)} \, \tilde \phi_4(\tau_1,\tau_2) \, u_{\Lambda_b}(v,s) \,, 
\end{align} 
for the chiral-odd Dirac structures,
and 
\begin{align}
     \epsilon^{abc} \, \langle 0| \left(  u^a(\tau_1 n) \, 
  C \gamma_5  \,  d^b(\tau_2 n) \right) h_v^c(0) |\Lambda_b(v,s) \rangle 
  &= f_{\Lambda_b}^{(1)} \, \tilde \phi_3^s(\tau_1,\tau_2) \, u_{\Lambda_b}(v,s) \,,
   \nonumber \\[0.2em]
   \epsilon^{abc} \, \langle 0| \left(  u^a(\tau_1 n) \, 
  C \gamma_5 \, i\sigma_{\mu\nu} \bar n^\mu n^\nu  \,  d^b(\tau_2 n) \right) h_v^c(0) |\Lambda_b(v,s) \rangle 
  &= 2f_{\Lambda_b}^{(1)} \, \tilde \phi_3^\sigma(\tau_1,\tau_2) \, u_{\Lambda_b}(v,s) \,,
\end{align}
for the chiral-even ones.
Here, gauge links between the different space-time points are understood 
\emph{implicitly}, i.e.\ 
$$
 u(\tau_1 n) \to [0,\tau_1 n] \, u(\tau_1 n)  \,, \qquad 
 d(\tau_2 n) \to [0,\tau_2 n] \, d(\tau_2 n) \,,
$$
where $[0,\tau_i n]$ is a straight QCD Wilson line connecting the points $\tau_i n^\mu$ to the 
location of the quasi-static heavy-quark field $h_v(0)$ at the origin.
The normalization factors $f_{\Lambda_b}^{(1,2)}$ are defined by the matrix elements of the corresponding \emph{local} operators,
\begin{align} 
 \epsilon^{abc} \, \langle 0| \left(  u^a(0) \, 
  C \gamma_5 \gamma^\mu  \,  d^b(0) \right) h_v^c(0) |\Lambda_b(v,s) \rangle 
  &= f_{\Lambda_b}^{(2)} \, v^\mu \, u_{\Lambda_b}(v,s) \,,
\cr 
\epsilon^{abc} \, \langle 0| \left(  u^a(0) \, 
  C \gamma_5  \,  d^b(0) \right) h_v^c(0) |\Lambda_b(v,s) \rangle 
  &= f_{\Lambda_b}^{(1)} \, u_{\Lambda_b}(v,s) \,.
\label{eq:fdef}
\end{align}
Notice that the parameters $f_{\Lambda_b}^{(1,2)}$ are renormalization-scale dependent, since the defining currents are not conserved in HQET. Moreover, as we will verify explicitly below, the local limit ($\tau_i \to 0$) of the LCDAs  is non-trivial.

Later, we will also consider the LCDAs in momentum space which are defined by the Fourier transform
\begin{align}
    \phi_i(\omega_1,\omega_2) &= \int\limits_{-\infty-i\epsilon}^{\infty - i\epsilon}  \frac{d\tau_1}{2\pi} \, 
    \int\limits_{-\infty-i\epsilon}^{\infty-i\epsilon} \frac{d\tau_2}{2\pi} \, e^{i \omega_1 \tau_1 + i \omega_2\tau_2} \, \tilde \phi_i(\tau_1,\tau_2) \,.
    \label{eq:momLCDA}
\end{align}
with an appropriate $i\epsilon$ description, 
which can also be understood as an inverse Laplace transform with respect to the variables $s_i = i \tau_i $.
Notice that the analytic properties of the functions $\tilde \phi_i(\tau_1,\tau_2)$ in HQET ensure that the momentum-space LCDAs only have support for $\omega_i\geq 0$.

\section{Short-distance expansion}

\label{sec:ope}

For short distances, $\tau_{1,2} \ll 1/\Lambda_{\rm QCD}$, the light-ray operators that define the above LCDAs can be expanded into a tower of local operators of increasing canonical mass dimension. For the corresponding 2-particle operators that define the $B$-meson LCDAs, this has already been discussed in \cite{Kawamura:2008vq} for $B_{u,d}$ mesons with massless spectator quarks and in \cite{Feldmann:2023aml} for $B_s$ mesons with a massive spectator quark.
In our case, we define 
\begin{align}
    {\cal O}_\Gamma(\tau_1,\tau_2) & \equiv 
    \epsilon^{abc} \left(  u^a(\tau_1 n) \, 
  C \gamma_5 \, \Gamma \,  d^b(\tau_2 n) \right) h_v^c(0)
  \cr 
  & = \sum_{i=0}^\infty \sum_P \sum_{k=1}^{K_i} c_{P,k}^{(i+9/2)}(\tau_1,\tau_2) \,
  {\cal O}_{P\Gamma,k}^{(i+9/2)}(0)
  \label{eq:ope}
\end{align}
for a generic Dirac matrix $\Gamma$ between the light-quark fields. 
For massless quarks, chiral-even and -odd structures would not mix into each other
(which would no longer be the case if one takes into account explicit masses $m_u=m_d \neq 0$, see \cite{Feldmann:2023aml,DV}). 
The possible Dirac structures for the local operators on the right-hand side are then restricted by Lorentz invariance to the projections 
$$
P\Gamma \ : \ 
 \Gamma_{++} = \frac{\slashed n \slashed {\bar n}}{4} \, \Gamma \, \frac{\slashed {\bar n} \slashed {n}}{4} \,, \quad 
 \Gamma_{+-} = \frac{\slashed n \slashed {\bar n}}{4} \, \Gamma \, \frac{\slashed n \slashed {\bar n}}{4} \,, 
 \quad 
 \Gamma_{-+} = \frac{\slashed {\bar n} \slashed {n}}{4}  \, \Gamma \, \frac{\slashed {\bar n} \slashed { n}}{4}\,, 
 \quad 
 \Gamma_{--} = \frac{\slashed {\bar n} \slashed {n}}{4}  \, \Gamma \, \frac{\slashed {n} \slashed {\bar n}}{4} \,.
$$
Specifically, for the operator defining the LCDA $\tilde \phi_2(\tau_1,\tau_2)$,
we have $\Gamma=\slashed n$, and the only non-vanishing Dirac structures 
involve the projection $\Gamma_{++} = \slashed n$, such that the short-distance expansion reads
\begin{align}
  \epsilon^{abc} \left(  u^a(\tau_1 n) \, 
  C \gamma_5 \, \slashed n \,  d^b(\tau_2 n) \right) h_v^c(0)  
 &=
  \epsilon^{abc} \Bigg( u^a(0) \, C\gamma_5  \, \slashed n \, {\cal D}(\tau_1,\tau_2) \, d^b(0) \Bigg) h_v^c(0)  \,,
\end{align}
\newpage
with
\begin{align}
    {\cal D}(\tau_1,\tau_2) &= c_{++}^{(9/2)}(\tau_1,\tau_2) 
  \cr 
  &  {} +
  c_{++,1}^{(11/2)}(\tau_1,\tau_2) \,  (i n \cdot \overleftarrow D) \, 
  + c_{++,2}^{(11/2)}(\tau_1,\tau_2) \, (i n \cdot \overrightarrow D) 
  \cr 
  &
  {} + c_{++,3}^{(11/2)}(\tau_1,\tau_2) \,  (i v \cdot \overleftarrow D) 
  + c_{++,4}^{(11/2)}(\tau_1,\tau_2) \, (i v \cdot \overrightarrow D) + \ldots 
\end{align}
As a consequence of isospin symmetry, we have 
$$
 c_{++}^{(9/2)}(\tau_1,\tau_2) = c_{++}^{(9/2)}(\tau_2,\tau_1) \,, \quad 
 c_{++,1}^{(11/2)}(\tau_1,\tau_2) = c_{++,2}^{(11/2)}(\tau_2,\tau_1) \,, 
 \quad 
  c_{++,3}^{(11/2)}(\tau_1,\tau_2) = c_{++,4}^{(11/2)}(\tau_2,\tau_1) \,, 
$$
Similarly, for the operator defining the LCDA $\tilde \phi_4(\tau_1,\tau_2)$,
we have $\Gamma=\slashed {\bar n}$, and the only non-vanishing Dirac structures 
involve the projection $\Gamma_{--} = \slashed {\bar n}$.
The short-distance expansion up to dimension-11/2 is thus given in terms of
the Wilson coefficients
$$
 c_{--}^{(9/2)}(\tau_1,\tau_2) = c_{--}^{(9/2)}(\tau_2,\tau_1) \,, \quad 
 c_{--,1}^{(11/2)}(\tau_1,\tau_2) = c_{--,2}^{(11/2)}(\tau_2,\tau_1) \,, 
 \quad 
  c_{--,3}^{(11/2)}(\tau_1,\tau_2) = c_{--,4}^{(11/2)}(\tau_2,\tau_1) \,.
$$

For the operator defining the chiral-even LCDA $\tilde \phi_3^s(\tau_1,\tau_2)$
we have $\Gamma=1$, and thus the projections $\Gamma_{+-} = \frac{\slashed n \slashed{\bar n}}{4}$ and $\Gamma_{-+} = \frac{\slashed {\bar n} \slashed n}{4}$
contribute, such that the short-distance expansion up to dimension-11/2 is defined in terms of the Wilson coefficients
$$
 c_{+-}^{(9/2)}(\tau_1,\tau_2) = c_{-+}^{(9/2)}(\tau_2,\tau_1) \,, 
$$ 
and 
$$
 c_{+-,1}(\tau_1,\tau_2) = c_{-+,2}(\tau_2,\tau_1) \,, \qquad 
 c_{-+,1}(\tau_1,\tau_2) = c_{+-,2}(\tau_2,\tau_1) \,,
$$
and 
$$
 c_{+-,3}(\tau_1,\tau_2) = c_{-+,4}(\tau_2,\tau_1) \,, \qquad 
 c_{-+,3}(\tau_1,\tau_2) = c_{+-,4}(\tau_2,\tau_1)
$$
The same coefficients appear in the expansion of the operator defining $\tilde \phi_3^\sigma(\tau_1,\tau_2)$.
The NLO matching for these Wilson coefficients will be performed in the next section.

\section{Calculation of the matching coefficients}
\label{sec:fixed-order}

We start with the tree-level matching, which simply follows from the Taylor expansion of the light-quark fields,
$$
 q(\tau n) = q(0) - i \tau \left[ (i n \cdot \partial) q \right](0) + \ldots 
$$
from which we read off 
\begin{align}
    c_{P}^{(9/2)}(\tau_1,\tau_2)  &= 1 + {\cal O}(\alpha_s) \,, 
    \cr 
    c_{P,1}^{(11/2)}(\tau_1,\tau_2) = c_{P,2}^{(11/2)}(\tau_2,\tau_1) &= - i \tau_1 + {\cal O}(\alpha_s) \,, 
    \cr 
    c_{P,3}^{(11/2)}(\tau_1,\tau_2) = c_{P,4}^{(11/2)}(\tau_2,\tau_1) &= {\cal O}(\alpha_s) \,. 
\end{align}

\begin{figure}[t!]
\begin{center}
\begin{tabular}{ccc}
 a1) & a2) & d)
\\ 
 \includegraphics[width=0.27\linewidth]{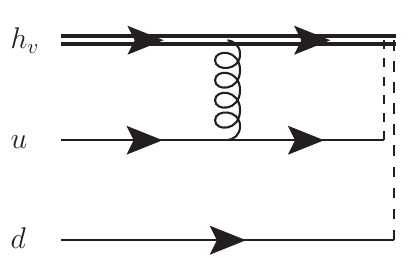}
& 
 \includegraphics[width=0.27\linewidth]{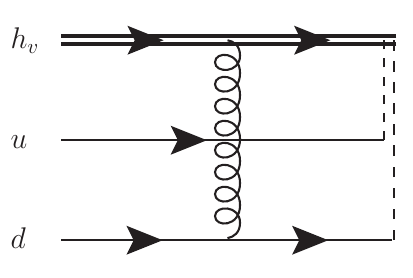}
 &
 \includegraphics[width=0.27\linewidth]{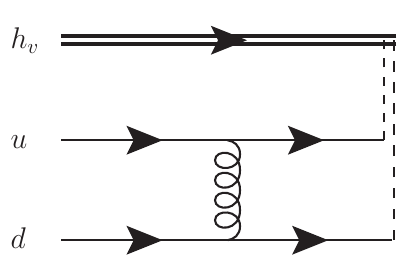} 
 \\[1.5em]
  b1) & c1) & c3)
\\ 
 \includegraphics[width=0.27\linewidth]{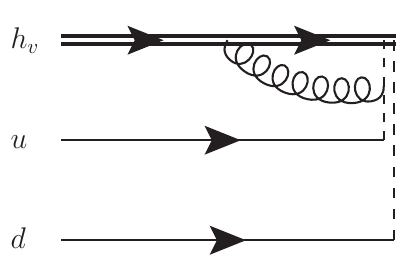}
& 
 \includegraphics[width=0.27\linewidth]{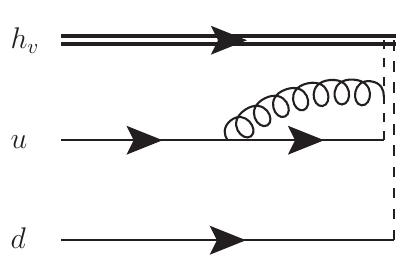}
 &
 \includegraphics[width=0.27\linewidth]{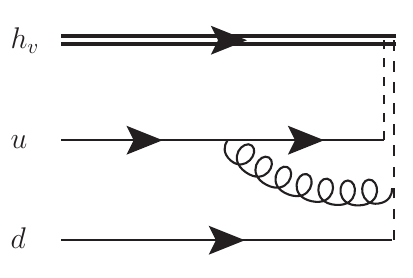} 
 \\[1.5em]
  b2) & c2) & c4)
\\ 
 \includegraphics[width=0.27\linewidth]{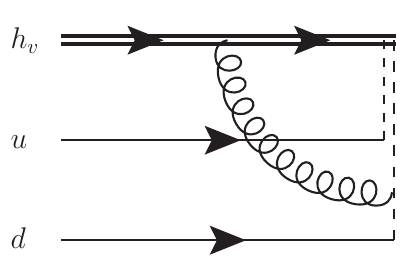}
& 
 \includegraphics[width=0.27\linewidth]{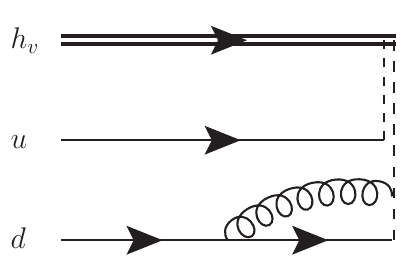}
 &
 \includegraphics[width=0.27\linewidth]{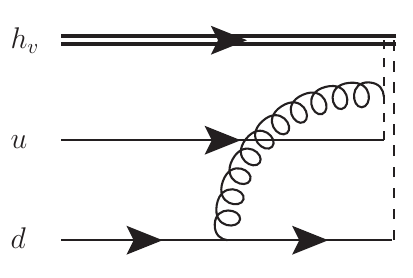} 
 \end{tabular}
\end{center}
\caption{\label{fig:feyn} Feynman diagrams contributing to the matching calculation at NLO. Here, the double line represents the heavy quark, and the dashed lines indicate the two Wilson lines connected to the light quarks.}
\end{figure}

\subsection{One-loop diagrams}

At next-to-leading order (NLO) in the strong coupling, we perform the matching calculation by comparing the matrix elements of the operators on the left- and right-hand side of Eq.~(\ref{eq:ope}) between \emph{partonic} on-shell states and the vacuum. We then have to consider the Feynman diagrams in Fig.~\ref{fig:feyn} that determine the ${\cal O}(\alpha_s)$ corrections to the light-ray operators ${\cal O}_\Gamma(\tau_1,\tau_2)$. (The same diagrams appear in the derivation of the evolution equations of the baryon LCDAs that govern the renormalization-scale dependence in HQET \cite{Ball:2008fw}, see also \cite{Bell:2013tfa,Braun:2014npa}.)
These Feynman diagrams can be evaluated in a straight-forward manner, following -- for instance -- the approach outlined for the mesonic case in \cite{Feldmann:2023aml}. In particular, we use the same strategy to subtract the NLO corrections to the \emph{local} operators on the right-hand side of Eq.~(\ref{eq:ope}) by first calculating the Feynman diagrams 
in momentum space, and then subtracting the local contributions when performing the Fourier transform to position space. Schematically,
$$
  \tilde I_{\rm subtr.}(\tau_1,\tau_2) = \int_0^\infty d\omega_1 \, \int_0^\infty d\omega_2 \left( e^{-i \omega_1 \tau_1 - i \omega_2 \tau_2} - 1 + i \omega_1 \tau_1 + i \omega_2 \tau_2 - \ldots \right) I(\omega_1,\omega_2)
\,. $$
In this way, the IR divergences from small values of $\omega_{1,2}$ drop out to the considered order of the local expansion, while the mismatch in the UV region ($\omega_{1,2} \gg 1/\tau_{1,2}$) becomes manifest when evaluating $I(\omega_1,\omega_2)$ in $D=4-2\epsilon$ dimensions.
Let us discuss the calculation of the different types of diagrams in turn. 

\subsubsection*{a1) and a2): Coupling between the heavy quark and one light quark}

The diagrams a1) and a2) are similar to the mesonic case, with the only difference that one additional light quark (including its Wilson line) acts as a genuine spectator that is unaffected by the gluon exchange.
We therefore have (all 1-loop integrals in units of $\frac{\alpha_s C_F}{4\pi}$)
\begin{align}
  I_{a1}^\Gamma(\omega_1,\omega_2) &= I_a^\Gamma(\omega_1) \, \delta(\omega_2- n\cdot k_2) \,, 
\end{align}
where here and in the following $k_{1,2}^\mu$ denote  the on-shell momenta of the external light quarks.
The Feynman integral $I_a^\Gamma(\omega_1, n\cdot k_1)$ can be obtained from the 
corresponding expression in Eq.~(A.9) in \cite{Feldmann:2023aml}
\begin{align}
    I_a^\Gamma(\omega_1) &=  \frac12 \epsilon^{abc} \,\frac{\Gamma(1+\epsilon)}{\omega_1} \left( \frac{\mu^2 e^{\gamma_E}}{\omega_1^2} \right)^\epsilon 
    \cr 
   & {} \times u^a(k_1) \, C\gamma_5  \Bigg\{ \left( 
    2+(1+2\epsilon) \, \frac{n\cdot k_1}{\omega_1} + (4+2\epsilon) \, 
    \frac{v\cdot k_1}{\omega_1} \right) \frac{\slashed n \slashed{\bar n}}{4} 
    \cr 
    & \qquad \qquad \qquad  {} + \left( - 2 -(1+2\epsilon) \, \frac{n\cdot k_1}{\omega_1} + (2-2\epsilon) \, \frac{v\cdot k_1}{\omega_1} \right) \frac{\slashed {\bar n} \slashed n}{4} 
    \cr 
    & \qquad \qquad \qquad  {} + {\cal O}(\omega_1^{-2}) \Bigg\} 
    \Gamma \, d^b(k_2) \, u^c(v) \,,
\end{align}
with more or less obvious replacements for the external quark spinors, and taking into account a different overall color factor ($C_F/2$ in the diquark channel, compared to $C_F$ in the meson channel).\footnote{We remind the reader that we consider massless light quarks in this work. Taking into account quark masses, we also encounter some relative-sign flips for the mass terms, compared to the meson case, see \cite{DV}.}  
\newpage
Performing the matching on the level of the Fourier transform as outlined above, we obtain the same kind of integrals as in the mesonic case, leading to
\begin{align}
    \tilde I_{a1}^\Gamma(\tau_1,\tau_2)_{\rm subtr.} &= 
    \int_0^\infty d\omega_1 \left( e^{-i \omega_1\tau_1} - 1 + i \omega_1\tau_1 
    + i \tau_2 \, n \cdot k_2 \left( e^{-i\omega_1 \tau_1} - 1 \right) + \ldots \right) 
    I_{a}^\Gamma(\omega_1) 
    \cr 
    &=  \frac12 \epsilon^{abc} \, u^a(k_1) \, C\gamma_5  \Bigg\{ 
    \cr & \qquad \qquad 
    \left( \left( \frac{1}{2\epsilon} + L_1 \right) i\tau_1 \, n\cdot k_1 
    + \left( \frac{2}{\epsilon} + 4 L_1 - 3 \right) i\tau_1 \, v\cdot k_1 
    \right. 
    \cr & \qquad \qquad \qquad \left. {}
   + \left( 
    -\frac{1}{\epsilon} - 2 L_1 \right) (1-i\tau_2 \, n \cdot k_2) 
    \right) 
    \frac{\slashed n \slashed {\bar n}}{4} 
    \cr 
    & \qquad \quad \ {} +  \left( \left( -\frac{1}{2\epsilon} - L_1 \right) i\tau_1 \, n\cdot k_1 
    + \left( \frac{1}{\epsilon} + 2 L_1 - 3 \right) i\tau_1 \, v\cdot k_1 
    \right. 
    \cr & \qquad \qquad \qquad \left. {}
   + \left( 
    \frac{1}{\epsilon} + 2 L_1 \right) (1-i\tau_2 \, n \cdot k_2) 
    \right) 
    \frac{\slashed {\bar n} \slashed {n}}{4} 
     \Bigg\}\Gamma \, d^b(k_2) \, u^c(v) \,,
\end{align}
where terms of order $\tau_i^2$ have been dropped. 
Here we defined the abbreviation
$$
 L_i = \ln (i \tau_i \, \mu e^{\gamma_E}) \,,
$$
and our definition of $\epsilon$ already includes the pre-factors for $\overline{\rm MS}$ subtraction. 
From this, one can read off the contributions to the Wilson coefficients in a straight-forward manner. In the $\overline{\rm MS}$ scheme we obtain
\begin{align}
    c_{++}^{(9/2)}(\tau_1,\tau_2) \big|_{a_1} 
    = c_{+-}^{(9/2)}(\tau_1,\tau_2) \big|_{a_1} &= - \frac{\alpha_s C_F}{4\pi} \left( L_1 \right)  \,, 
    \cr 
    c_{-+}^{(9/2)}(\tau_1,\tau_2) \big|_{a_1} 
    = c_{--}^{(9/2)}(\tau_1,\tau_2) \big|_{a_1} &=  - \frac{\alpha_s C_F}{4\pi} \left(  - L_1 \right)  \,,
\end{align}
and 
\begin{align}
    c_{++,1}^{(11/2)}(\tau_1,\tau_2) \big|_{a_1} 
    = c_{+-,1}^{(11/2)}(\tau_1,\tau_2) \big|_{a_1} &= i\tau_1 \, \frac{\alpha_s C_F}{4\pi} \left( \frac{L_1}{2} \right)  \,, 
        \cr 
    c_{-+,1}^{(11/2)}(\tau_1,\tau_2) \big|_{a_1} 
    = c_{--,1}^{(11/2)}(\tau_1,\tau_2) \big|_{a_1} &=  i\tau_1 \, \frac{\alpha_s C_F}{4\pi} \left( -\frac{L_1}{2} \right) \,,
\end{align}
and 
\begin{align}
    c_{++,2}^{(11/2)}(\tau_1,\tau_2) \big|_{a_1} 
    = c_{+-,2}^{(11/2)}(\tau_1,\tau_2) \big|_{a_1} &= i\tau_2 \, \frac{\alpha_s C_F}{4\pi} \left( L_1 \right)  \,, 
        \cr 
    c_{-+,2}^{(11/2)}(\tau_1,\tau_2) \big|_{a_1} 
    = c_{--,2}^{(11/2)}(\tau_1,\tau_2) \big|_{a_1} &=  i\tau_2 \, \frac{\alpha_s C_F}{4\pi} \left( - L_1 \right) \,,
\end{align}
and 
\begin{align}
    c_{++,3}^{(11/2)}(\tau_1,\tau_2) \big|_{a_1} 
    = c_{+-,3}^{(11/2)}(\tau_1,\tau_2) \big|_{a_1} &= i\tau_1 \, \frac{\alpha_s C_F}{4\pi} \left( 2 L_1 - \frac32 \right)  \,, 
        \cr 
    c_{-+,3}^{(11/2)}(\tau_1,\tau_2) \big|_{a_1} 
    = c_{--,3}^{(11/2)}(\tau_1,\tau_2) \big|_{a_1} &=  i\tau_1 \, \frac{\alpha_s C_F}{4\pi} \left( L_1- \frac32 \right) \,,
\end{align}
and 
\begin{align}
    c_{P,4}(\tau_1,\tau_2) \big|_{a1} & = 0 \,. 
\end{align}
The computation for the diagram a2) goes in a completely analogous way. The resulting contributions to the Wilson coefficients can be inferred from isospin symmetry in an obvious way.

\subsubsection*{b1) and b2): Coupling between the heavy quark and a Wilson line}

The diagrams b1) and b2) are particularly simple because the exchanged gluon does not resolve the Dirac structure between the light-quark fields. As a consequence, we can again derive the result from the corresponding calculation for the mesonic case, see Eq.~(A.11) in \cite{Feldmann:2023aml}. As before, we thus have
$$
 I_{b1}^\Gamma(\omega_1,\omega_2) = I_b^\Gamma(\omega_1) \, \delta(\omega_2-n\cdot k_2) \,,
$$
with 
\begin{align}
    I_b^\Gamma(\omega_1) &=  
    2\Gamma(\epsilon)
    \, \int_0^\infty dk \left( \frac{\mu^2 e^{\gamma_E}}{k^2} \right)^\epsilon \frac{\delta(\omega_1-n\cdot k_1-k) - \delta(\omega_1-n\cdot k_1)}{k}
    \cr 
   & \qquad {}  \times \frac12 \, \epsilon^{abc} \, u^a(k_1) \, C\gamma_5  \,  \Gamma \, d^b(k_2) \, u^c(v) \,.
\end{align}
For the b-type diagrams the local subtraction terms is absent, and the Fourier transform to position space simply yields 
\begin{align}
  \tilde I_{b1}^\Gamma(\tau_1,\tau_2) &= \left(1- i \tau_1 n\cdot k_1 - i \tau_2 n\cdot k_2 + \ldots \right) 
  \left( -\frac{1}{\epsilon^2} - \frac{2L_1}{\epsilon}- 2 L_1^2 - \frac{5\pi^2}{12} \right) 
  \cr 
    & \qquad {} \times \frac12 \, \epsilon^{abc} \, u^a(k_1) \, C\gamma_5  \,  \Gamma \, d^b(k_2) \, u^c(v) 
  \,,
\end{align}
and analogously for the diagram (b2).
From this we read off the contributions to the Wilson coefficients,
\begin{align}
    c_{P}^{(9/2)}(\tau_1,\tau_2)\big|_{b1+b2} = c_{P;1,2}^{(11/2)}(\tau_1,\tau_2)\big|_{b1+b2} &= - \frac{\alpha_s C_F}{4\pi} \left( L_1^2+L_2^2 + \frac{5\pi^2}{12} \right) \,,
    \cr 
    c_{P;3,4}^{(11/2)}(\tau_1,\tau_2)\big|_{b1+b2} &=0 \,. 
\end{align}

The remaining diagrams (c1-c4,d) do not contribute to the radiative tail, since the heavy-quark propagator is not involved. This is also reflected by the analytic properties of the resulting Feynman integrals, which constrain the
integration regions in the Fourier integral to values $\omega_i \leq n \cdot k_i$, where the multipole expansion of the Fourier factor and the integration commute.

\subsection{Summary of NLO results}

Combining the results from the individual diagrams,
we obtain for the Wilson coefficients at dimension-9/2:
\begin{align}
    c_{++}^{(9/2)}(\tau_1,\tau_2) &= 1 - \frac{\alpha_s C_F}{4\pi} 
    \left( L_1^2 + L_1 + L_2^2 + L_2 + \frac{5\pi^2}{12}\right) + {\cal O}(\alpha_s^2) \,, 
    \cr 
    c_{--}^{(9/2)}(\tau_1,\tau_2) &= 1- \frac{\alpha_s C_F}{4\pi} \left( 
    L_1^2-L_1 + L_2^2-L_2 + \frac{5\pi^2}{12} \right) + {\cal O}(\alpha_s^2) \,,
    \cr 
    c_{+-}^{(9/2)}(\tau_1,\tau_2) = c_{-+}^{(9/2)}(\tau_2,\tau_1) &= 
    1 - \frac{\alpha_s C_F}{4\pi} \left( L_1^2+ L_1 + L_2^2 - L_2 + \frac{5\pi^2}{12} \right) + {\cal O}(\alpha_s^2) \,.
\end{align}
The matching coefficients at dimension-11/2 follow as 
\begin{align}
    c_{++,1}^{(11/2)}(\tau_1,\tau_2) = c_{++,2}^{(11/2)}(\tau_2,\tau_1) &=
    -i\tau_1 \left( 1 - \frac{\alpha_s C_F}{4\pi} \left( 
    L_1^2+ \frac{L_1}{2} + L_2^2 + L_2 + \frac{5\pi^2}{12} \right) + {\cal O}(\alpha_s^2) \right)  \,, 
    \cr 
     c_{--,1}^{(11/2)}(\tau_1,\tau_2) = c_{--,2}^{(11/2)}(\tau_2,\tau_1) &=
    -i\tau_1 \left( 1 - \frac{\alpha_s C_F}{4\pi} \left( 
    L_1^2 - \frac{L_1}{2} + L_2^2 - L_2 + \frac{5\pi^2}{12} \right) + {\cal O}(\alpha_s^2) \right)  \,, 
    \cr 
    c_{+-,1}^{(11/2)}(\tau_1,\tau_2) = c_{-+,2}^{(11/2)}(\tau_2,\tau_1)  &=
    -i\tau_1 \left( 1 - \frac{\alpha_s C_F}{4\pi} \left( 
    L_1^2 + \frac{L_1}{2} + L_2^2 - L_2 + \frac{5\pi^2}{12} \right) + {\cal O}(\alpha_s^2) \right)  \,, 
    \cr 
    c_{-+,1}^{(11/2)}(\tau_1,\tau_2) = c_{+-,2}^{(11/2)}(\tau_2,\tau_1)  &=
    -i\tau_1 \left( 1 - \frac{\alpha_s C_F}{4\pi} \left( 
    L_1^2 - \frac{L_1}{2} + L_2^2 + L_2 + \frac{5\pi^2}{12} \right) + {\cal O}(\alpha_s^2) \right) 
\end{align}
and 
\begin{align}
    c_{++,3}^{(11/2)}(\tau_1,\tau_2) = c_{++,4}^{(11/2)}(\tau_2,\tau_1) &=
    -i \tau_1 \left( - \frac{\alpha_s C_F}{4\pi} \left( 2 L_1 - \frac32 \right) + {\cal O}(\alpha_s^2) \right) \,,
    \cr 
     c_{--,3}^{(11/2)}(\tau_1,\tau_2) = c_{--,4}^{(11/2)}(\tau_2,\tau_1) &=
    -i \tau_1 \left( - \frac{\alpha_s C_F}{4\pi} \left(  L_1 - \frac32 \right) + {\cal O}(\alpha_s^2) \right) \,,
  \cr 
  c_{+-,3}^{(11/2)}(\tau_1,\tau_2) = c_{-+,4}^{(11/2)}(\tau_2,\tau_1) &=
    -i \tau_1 \left( - \frac{\alpha_s C_F}{4\pi} \left( 2 L_1 - \frac32 \right) + {\cal O}(\alpha_s^2) \right) \,,
    \cr  
     c_{-+,3}^{(11/2)}(\tau_1,\tau_2) = c_{+-,4}^{(11/2)}(\tau_2,\tau_1) &=
    -i \tau_1 \left( - \frac{\alpha_s C_F}{4\pi} \left( L_1 - \frac32 \right) + {\cal O}(\alpha_s^2) \right) \,. 
\end{align}
These expressions represent one of the main results of this paper. 
In the next paragraph, we will provide the corresponding 
results for the LCDAs
that arise from taking hadronic matrix elements of the local operators in Eq.~(\ref{eq:ope}).

\section{Hadronic matrix elements and LCDAs}

\label{sec:lcdas}

In order to translate the short-distance expansion of the baryonic light-ray operators to the corresponding asymptotic behaviour of the LCDAs, we need to parametrize the relevant hadronic matrix elements of the local operators on the r.h.s.\ of Eq.~(\ref{eq:ope}). From Lorentz invariance and HQET spin symmetry, we find the expression
\begin{align}
    \epsilon^{abc} \, 
    \langle 0| \left( (u^a C\gamma_5)_\alpha d^b_\beta \right) h_{v,\gamma}^c |\Lambda_b(v,s)\rangle 
    &= \frac14 \left[M^{(9/2)}(v) \right]_{\beta\alpha} \, u_{\gamma}(v,s)
\end{align}
for a generic dimension-9/2 operator, with
\begin{align}
    M^{(9/2)}(v) &= f_{\Lambda_b}^{(1)} + f_{\Lambda_b}^{(2)} \, \slashed v \,,
\end{align}
implementing Eq.~(\ref{eq:fdef}).
Similarly, for the dimension-11/2 operators we define 
\begin{align}
    \epsilon^{abc} \, 
    \langle 0| \left( (u^a C\gamma_5)_\alpha (i \overleftarrow D_\mu) \, d^b_\beta \right) h_{v,\gamma}^c |\Lambda_b(v,s)\rangle 
    &= \frac14 \left[M_\mu^{(11/2)}(v) \right]_{\beta\alpha} \, u_{\gamma}(v,s) \,,
\end{align}
with 
\begin{align} 
M_\mu^{(11/2)}(v) &=  f_{\Lambda_b}^{(2)} \left( C \, \gamma_\mu + D \, \slashed v \, v_\mu \right)
+  f_{\Lambda_b}^{(1)} \left( E \, \slashed v \gamma_\mu + F \, \gamma_\mu \slashed v \right) \,.
\end{align}
Using the equations of motion for the light-quark fields, we obtain
\begin{align}
    0 &= \frac{1}{4} \, {\rm tr}[M_\mu^{(11/2)} \gamma^\mu] =
      f_{\Lambda_b}^{(2)}\left( 4 C+ D \right) \,, \cr 
    0 &= \frac14 \, {\rm tr}[M_\mu^{(11/2)} \gamma^\mu \slashed v] = 
    f_{\Lambda_b}^{(1)} \left( 4 E - 2 F \right) \,.
\end{align}
On the other hand, from the definition of the HQET mass parameter  $\bar\Lambda = M_{\Lambda_b} - m_b$, 
\begin{align}
    i v\cdot\partial \langle 0| \ldots | \Lambda_b(v,s) \rangle = \bar \Lambda \, \langle 0 |\ldots |\Lambda_b(v,s)\rangle \,, 
\end{align}
we obtain the relations
\begin{align}
    \bar \Lambda \, f_{\Lambda_b}^{(1)} &= 2 \cdot \frac{1}{4} \, {\rm tr} \left[v \cdot M^{(11/2)}\right] = 2 \,  f_{\Lambda_b}^{(1)} \, (E+F) \,, 
    \cr 
     \bar \Lambda \, f_{\Lambda_b}^{(2)} &= 2 \cdot \frac{1}{4} \, {\rm tr} \left[v \cdot M^{(11/2)} \slashed v \right] = 2 \,  f_{\Lambda_b}^{(2)} \, (C+D )\,,
\end{align}
where we have used isospin symmetry to relate the matrix elements with $(iv\cdot D)$ acting on the light-quark fields to the left or right
(where in the latter case the coefficients $E$ and $F$ change their role).
The four equations are then solved by 
\begin{align}
    C = - \frac{\bar\Lambda }{6} \,, \quad 
    D = \frac{2\bar\Lambda }{3} \,, \qquad
    E = \frac{\bar\Lambda }{6} \,, \quad 
    F = \frac{\bar\Lambda }{3} \,.
\end{align}

\subsection{Short-distance expansion for LCDAs}

With these relations, the short-distance epxansion of the different 3-particle LCDAs can be calculated in a straightforward manner, leading to
\begin{align}
\tilde \phi_2(\tau_1,\tau_2) &=
c_{++}^{(9/2)}(\tau_1,\tau_2) 
+ D \left( c_{++,1}^{(11/2)}(\tau_1,\tau_2) + c_{++,2}^{(11/2)}(\tau_1,\tau_2) \right) 
\cr & \qquad {} 
+ (C+D) \left( c_{++,3}^{(11/2)}(\tau_1,\tau_2)+c_{++,4}^{(11/2)}(\tau_1,\tau_2) \right) 
\cr 
\Rightarrow \quad \tilde \phi_2(\tau_1,\tau_2) &=
\left(1-i(\tau_1+\tau_2) \, \frac{2\bar\Lambda}{3} \right)
\left(1- \frac{\alpha_s C_F}{4\pi} \left( L_1^2+L_2^2 + L_1 + L_2 + \frac{5\pi^2}{12} \right) \right) 
\cr & \quad {} + i\bar \Lambda \, \frac{\alpha_s C_F}{4\pi} 
\left( \tau_1 \left( \frac{2 L_1}{3} - \frac34 \right) 
+ \tau_2 \left( \frac{2L_2}{3}- \frac34\right) \right)  + {\cal O}(\tau_i^2)
\,, 
\cr 
& \label{eq:phi2ope}
\\
\tilde \phi_4(\tau_1,\tau_2) &=
c_{--}^{(9/2)}(\tau_1,\tau_2) 
+ (2C+D) \left( c_{--,1}^{(11/2)}(\tau_1,\tau_2) + c_{--,2}^{(11/2)}(\tau_1,\tau_2) \right) 
\cr & \qquad {} 
+ (C+D) \left( c_{--,3}^{(11/2)}(\tau_1,\tau_2)+c_{--,4}^{(11/2)}(\tau_1,\tau_2) \right) 
\cr 
\Rightarrow \quad \tilde \phi_4(\tau_1,\tau_2)
&=
\left(1-i(\tau_1+\tau_2) \, \frac{\bar\Lambda}{3} \right)
\left(1- \frac{\alpha_s C_F}{4\pi} \left( L_1^2+L_2^2 - L_1 - L_2 + \frac{5\pi^2}{12} \right) \right) 
\cr & \quad {} + i\bar \Lambda \, \frac{\alpha_s C_F}{4\pi} 
\left( \tau_1 \left( \frac{2 L_1}{3} - \frac34 \right) 
+ \tau_2 \left( \frac{2L_2}{3}- \frac34\right) \right)  + {\cal O}(\tau_i^2)
\,,\cr 
\end{align}
\newline\newline
for the chiral-odd LCDAs, and
\begin{align}
    \tilde \phi_3^s(\tau_1,\tau_2) &=\frac{1}{2}\left(c_{+-}^{(9/2)}(\tau_1,\tau_2)+c_{-+}^{(9/2)}(\tau_1,\tau_2)\right) \cr
    &+F\left(c_{+-,1}^{(11/2)}(\tau_1,\tau_2)+c_{-+,2}^{(11/2)}(\tau_1,\tau_2)\right)+E\left(c_{-+,1}^{(11/2)}(\tau_1,\tau_2)+c_{+-,2}^{(11/2)}(\tau_1,\tau_2)\right) \cr
    &+\frac{1}{2}(E+F)\left(c_{+-,3}^{(11/2)}(\tau_1,\tau_2)+c_{+-,4}^{(11/2)}(\tau_1,\tau_2)+c_{-+,3}^{(11/2)}(\tau_1,\tau_2)+c_{-+,4}^{(11/2)}(\tau_1,\tau_2)\right) 
    \cr
    \Rightarrow \quad \tilde \phi_3^s(\tau_1,\tau_2) &=\left(1-i(\tau_{1}+\tau_{2})\frac{\bar{\Lambda}}{2}\right)\left(1-\frac{\alpha_{s}C_{F}}{4\pi}\left(L_{1}^2+L_{2}^2+\frac{5\pi^2}{12}\right)\right)\cr
     &+i \bar{\Lambda}\frac{\alpha_{s}C_{F}}{4\pi}\left(\tau_{1}\left(\frac{5}{6}L_{1}-\frac{1}{6}L_{2}-\frac{3}{4}\right)+\tau_{2}\left(\frac{5}{6}L_{2}-\frac{1}{6}L_{1}-\frac{3}{4}\right)\right)+ {\cal O}(\tau_i^2)
    \,, \cr 
    & \\
    \tilde \phi_3^{\sigma}(\tau_1,\tau_2) &=\frac{1}{2}\left(c_{+-}^{(9/2)}(\tau_1,\tau_2)-c_{-+}^{(9/2)}(\tau_1,\tau_2)\right) \cr
    &+F\left(c_{+-,1}^{(11/2)}(\tau_1,\tau_2)-c_{-+,2}^{(11/2)}(\tau_1,\tau_2)\right)+E\left(-c_{-+,1}^{(11/2)}(\tau_1,\tau_2)+c_{+-,2}^{(11/2)}(\tau_1,\tau_2)\right) \cr
    &+\frac{1}{2}(E+F)\left(c_{+-,3}^{(11/2)}(\tau_1,\tau_2)+c_{+-,4}^{(11/2)}(\tau_1,\tau_2)-c_{-+,3}^{(11/2)}(\tau_1,\tau_2)-c_{-+,4}^{(11/2)}(\tau_1,\tau_2)\right) 
    \cr
 \Rightarrow \quad \tilde \phi_3^\sigma(\tau_1,\tau_2)   &=\left(-i(\tau_{1}-\tau_{2})\frac{\bar{\Lambda}}{6}\right)\left(1-\frac{\alpha_{s}C_{F}}{4\pi}\left(L_{1}^2+L_{2}^2+\frac{5\pi^2}{12}\right)\right) \cr
    &+i \bar{\Lambda}\frac{\alpha_{s}C_{F}}{4\pi}\left((\tau_{1}+\tau_{2})\frac{1}{2}\left(L_{1}-L_{2}\right)\right)-\frac{\alpha_{s}C_{F}}{4\pi} \left(L_{1}-L_{2}\right)+ {\cal O}(\tau_i^2)
    \,.
\end{align}
for the chiral-even ones.

\subsection{Model-dependent extrapolation to large distances}

In order to illustrate our results, we propose 
a strategy to extrapolate the OPE result to large values of $\tau_i$.
By nature, this extrapolation is not unique but will require to model the long-distance behaviour of the LCDAs in some way. 
Let us explicitly focus on the LCDA $\tilde \phi_2(\tau_1,\tau_2)$ in the following (the other LCDAs can be treated in a similar way).
\begin{itemize}
\item To begin with, we simplify the discussion and consider the LCDA at equal light-cone distances as a function of the variable $s = i \tau_1 = i \tau_2$,
defining
$$
\tilde\phi(s) \equiv \tilde\phi_2(-is,-is) 
= \int_0^\infty d\omega \, \omega \, e^{-\omega s} \, \int_0^1 du \, \phi_2(u\omega,(1-u)\omega) \,,
$$
where the second relation follows from eq.~(\ref{eq:momLCDA}) and introducing the total and relative momenta
$$
\omega_1 = u \omega \,, \qquad \omega_2 = (1-u) \, \omega \,.
$$
\item The renormalization-group equation for the function $\tilde \phi(s)$ 
at 1-loop takes a similar form as in the $B$-meson case \cite{Kawamura:2010tj}, since the Brodsky-Lepage kernel does not contribute, see Appendix,
\begin{align}
  \frac{d\tilde\phi(s;\mu)}{d\ln\mu} 
  & = - \frac{\alpha_s C_F}{4\pi} \left( \Gamma_c^{(1)} \, L + \gamma_F^{(1)} \right) \tilde\phi(s;\mu) 
  \cr 
  & \quad {} 
  + \frac{\alpha_s C_F}{4\pi} \, \Gamma_c^{(1)} \int_0^1 dz \, \frac{z}{1-z} \left( \tilde \phi_2(-izs,-is;\mu) -\tilde \phi_2(-is,-is;\mu) \right) 
\,. 
\cr &
\label{eq:ourRGE}
\end{align} 
with $\Gamma_c^{(1)}=4$, $\gamma_F^{(1)}=2$, and  $L = \ln (s\mu e^{\gamma_E})$.
Notice that on the right-hand side still the function $\tilde \phi_2$ with non-equal arguments appears, and the convolution integral is only with respect to
one of the arguments.

\item 
At fixed order, we can replace the LCDA on the right-hand side by a ''tree-level'' model $\tilde f_2(s_1,s_2)$ with $\tilde f(s)=\tilde f_2(s,s)$, 
and integrate the differential equation  to obtain
\begin{align}
    \tilde \phi(s,\mu) &= 
    \left( 1 - \frac{\alpha_s C_F}{4\pi} \left( 
    \left( 4 L - 2 \ln \frac{\mu}{\mu_0(s)} \right) \ln \frac{\mu}{\mu_0(s)}
    + 2 \, \ln \frac{\mu}{\mu_0(s)} + {\rm const}_1(s) \right) \right) \tilde f(s) 
    \cr 
   & {} \quad 
    + \frac{\alpha_s C_F}{4\pi} \left( 4 \ln \frac{\mu}{\mu_0(s)} + 4 \, {\rm const}_2(s) \right) 
    \, \int_0^1 dz \, \frac{z}{1-z} \left( \tilde f_2(zs,s) -\tilde f_2(s,s) \right)
    \label{eq:interpol}
\end{align}
where the initial scale $\mu_0(s)$ and the two integration constants must not depend on $\mu$, but may still depend on $s$.
The fixed-order result for the radiative tail can then directly be reproduced for the choice
\begin{align}
    \mbox{case 1:} & \quad
    \mu_0(s) = e^{-\gamma_E}/s\,, \quad 
    {\rm const}_1 = \frac{5\pi^2}{12} \,, \quad 
    {\rm const}_2 = - \frac{9}{8} \,,
\end{align}
together with the constraints on the model function, $\tilde f(0)=1$ and $\tilde f'(0)=-4\bar\Lambda/3$.
This choice, however, has the drawback that for large values of $s$
the initial scale $\mu_0(s)$ tends to zero, which leads to a pathological behaviour of the resulting LCDA. 
To avoid this, we may consider a choice that yields a smooth transition to the shape of the original model for large values of $s$, 
\begin{align}
    \mbox{case 2:} & \quad
    \hat\mu_0(s) = \frac{1 + \hat\mu_F s}{s} \,, \quad 
    {\rm const}_1 = \frac{5\pi^2}{12} \,, \quad 
    {\rm const}_2 = - \frac{9}{8} \, \frac{1}{1+ \hat\mu_F s} \,,
    \label{eq:case2}
\end{align}
where here and in the following we use the abbreviation $\hat \mu= \mu e^{\gamma_E}$.
In this case, the initial scale approaches a constant value $\mu_F$ for large values of $s$, 
whereas the fixed-order expansion at small $s$ can be reproduced by re-adjusting the parameters in the tree-level model, such that
\begin{align}
 \tilde f(0) = 1 \,, \qquad 
 \tilde f'(0) = -\frac{4\bar \Lambda}{3} \left( 1 + \frac32 \, \frac{\alpha_s C_F}{4\pi} \, \frac{\hat\mu_F}{\bar \Lambda} \right) 
 \equiv -\frac{4 \bar\Lambda_{\rm eff}(\mu,\mu_F)}{3} \,.
\label{eq:fprimefix}
\end{align}
We emphasize that the choice for $\mu_0(s)$ in eq.~(\ref{eq:case2}) is in line with the general constraints on the analytic properties of the LCDAs in position space, i.e.\ $\tilde \phi(s=i\tau)$ is an analytic function 
for ${\rm Im}(\tau) <0$ (see also the discussion in \cite{Feldmann:2022uok}).
We also note that by extending the short-distance expansion to include dimension-13/2 operators,
one would have to consider  operators containing two covariant derivatives acting on the light-quark fields. 
The corresponding hadronic matrix elements would then constrain the second derivative $\tilde f''(0)$ in a similar way as in (\ref{eq:fprimefix}). 
In view of the overall theoretical uncertainties in exclusive $\Lambda_b$ decays, however, the inclusion of these or even
higher-order terms in the OPE is phenomenologically irrelevant at present.
\end{itemize}

\begin{figure}[t!b]
    \begin{center} 
    \fbox{\includegraphics[width=0.61\linewidth]{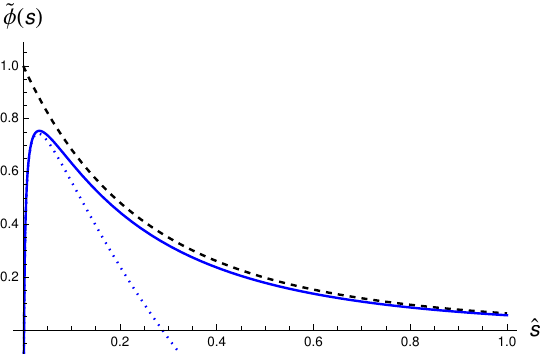}}
    \end{center}  
    \caption{Impact of the radiative tail on the light-cone distribution amplitude $\tilde \phi(s)$ evaluated at $s=i\tau_1=i\tau_2$ 
    as a function of the dimensionless variable $\hat s= s\omega_0$. The black dashed curve shows the
    input model in (\ref{eq:phi2tree}). 
    The blue dotted line shows the pure short-distance behaviour from eq.~(\ref{eq:phi2ope}). 
    The solid blue line refers to the interpolation between the two in eq.~(\ref{eq:interpol}). 
    For illustrative purposes, we took  $\alpha_s = 0.3$, $\mu e^{\gamma_E}= \mu_f e^{\gamma_E} = 2$~GeV
    and $\bar\Lambda = 0.6$~GeV as numerical input.
    With this choice, the parameter $\omega_0$ is fixed by eqs.~(\ref{eq:fprimefix},\,\ref{eq:phi2tree}) to $\omega_0 \simeq 0.23$~GeV.}
    \label{fig:phi2tilde}
\end{figure}
\newpage
For illustration, we consider a simple model (see e.g.\ \cite{Bell:2013tfa} and references therein),  
\begin{align}
\label{eq:phi2tree}
f_2(s_1,s_2) = \frac{1}{(1+\omega_0 s_1)^2 (1+\omega_0 s_2)^2} \quad \Rightarrow \quad 
f(s) &= \frac{1}{(1+\omega_0 s)^4} \,, \quad 
    f'(0) = -4\omega_0 \,, 
\end{align}
(alternative models or more sophisticated parameterisations can be chosen as well.)
For this model the $z$-integration in the extrapolation ansatz can be explicitly performed,
\begin{align}
    \int_0^1 dz \, \frac{z}{1-z} \left(\tilde f_2(zs,s)- \tilde f_2(s,s) \right) 
    = \frac{\ln(1+s\omega_0)}{(1+s\omega_0)^4} \,.
    \label{eq:modelconv}
\end{align}
In Fig.~\ref{fig:phi2tilde} we compare the OPE-improved model for $\tilde \phi(s)$ in eq.~(\ref{eq:interpol})
with the (''tree-level'') input model $f_2(s,s)$ on the one hand, and with the pure short-distance expansion in (\ref{eq:phi2ope}) on the other.
We observe the expected behaviour, in particular, the interpolation reproduces the shape of the original model input 
at values of $\hat s = s\omega_0 \gtrsim 1$ (corresponding to $|\tau|\gtrsim 1/\omega_0$), 
whereas for small values $\hat s \sim \omega_0/\mu \ll 1$ the behaviour of $\tilde \phi(s)$ is dictated by the short-distance expansion. 
Notice that for very small values $\hat s \ll \omega_0/\mu $, the fixed-order perturbative expansion breaks down, and one would 
have to further \emph{locally} resum large logarithms $\ln s \mu$, see the analogous discussion for the $B$-meson case in \cite{Feldmann:2014ika}.
\newpage
For later use, we write the final result for $\tilde \phi(s,\mu)$ in factorized form,
\begin{align}
    \tilde \phi(s,\mu) &= 
    \tilde K_1(s) \, s \tilde f(s) 
    + \tilde K_2(s) \, \left(1+\hat \mu_F s\right) \tilde f(s) 
    \cr 
    & \qquad {} + \tilde K_3(s) \, 
    \int_0^1 dz \, \frac{z}{1-z} \left( \tilde f_2(zs,s)-\tilde f_2(s,s) \right) \,,
    \label{eq:factorized}
\end{align}
with 
\begin{align} 
\tilde K_1(s) &=\frac1{s} \left(1 - \frac{\alpha_s C_F}{4\pi} 
    \left( 2 \ln^2 (\hat \mu s) +  2 \ln \frac{\hat\mu  s}{1+\hat\mu_F s} + \frac{5\pi^2}{12}  \right) \right) \,, 
  \cr  
  \tilde K_2(s) &= \frac{1}{1+\hat \mu_F s} \, 
  \frac{\alpha_s C_F}{2\pi} \, \ln^2 (1+\hat \mu_F s) \,, 
  \cr 
\tilde K_3(s)  & =
   \frac{\alpha_s C_F}{\pi}\left( \ln \frac{\hat \mu   s}{1+\hat \mu_F  s} -\frac9{8 \, (1+\hat\mu_F s)} \right) \,,
\end{align} 
The prefactors in the definition of the functions $\tilde K_i(s)$ have been chosen such that the inverse Laplace transform in momentum space takes a simple form, see below.
%


\subsection{Models with radiative tail in momentum space}

The momentum-space expression that corresponds to
the position-space LCDA at equal light-cone separations,
$\tilde \phi(s)$ is given by the inverse Laplace transform,
\begin{align}
    \phi(\omega) & \equiv 
    \omega \, \int_0^1 du \, \phi_2(u\omega,\bar u\omega) 
    = \int\limits_{-i\infty+\epsilon}^{i\infty +\epsilon} \, \frac{ds}{2\pi i} \,
    e^{\omega s} \, \tilde \phi(s) \,. 
\end{align}
With the factorized form in eq.~(\ref{eq:factorized}) 
this turns into  a convolution in momentum space,\footnote{The idea to implement the radiative tail via a convolution integral with a short-distance kernel has also been suggested for the modelling of the $B$-meson shape function appearing in the factorization approach to \emph{inclusive} $B$-meson decays in Ref.~\cite{Ligeti:2008ac}.}
\begin{align}
     \phi(\omega,\mu) &= 
    \int_0^\omega d\omega' \left( 
    K_1(\omega-\omega') \, \frac{df(\omega')}{d\omega'}
    + 
    K_2(\omega-\omega') \, 
    \left(1+\hat \mu_F \frac{d}{d\omega'} \right) f(\omega') \right) 
    \cr 
    & \qquad {} +  \int_0^\omega d\omega' \, K_3(\omega-\omega') 
    \cr 
    & \qquad {} \qquad {} \times 
    \left\{ f(\omega') 
    + \omega' \, 
    \int_0^1du \, \int_{u\omega'}^\infty d\omega_1'
    \left[ \frac{u\omega'}{\omega_1' \, (\omega_1'-u\omega')} \right]_+
    f_2(\omega_1',\bar u \omega')
    \right\} 
    \,.
    \cr &
    \label{eq:convolution}
\end{align}
For our model (\ref{eq:phi2tree})  we simply have 
\begin{align}
  f_2(\omega_1,\omega_2) &= 
  \frac{\omega_1 \omega_2}{\omega_0^4} \, e^{-(\omega_1+\omega_2)/\omega_0} \, 
  \,, \quad 
  f(\omega) = \frac{\omega^3}{6\omega_0^4} \, e^{-w/\omega_0} \,, 
  \quad 
  f'(\omega) = \left(\frac{3}\omega-\frac{1}{\omega_0} \right) f(\omega)
  \label{eq:fmodel}
\end{align}
and 
\begin{align} 
f_{\rm aux}(\omega') & \equiv 
f(\omega') + 
\omega' \, 
    \int_0^1du \, \int_{u\omega'}^\infty d\omega_1'
    \left[ \frac{u\omega'}{\omega_1' \, (\omega_1'-u\omega')} \right]_+
    f_2(\omega_1',\bar u \omega')
    \cr 
&= \left( - \ln \frac{\omega'}{\omega_0} - \gamma_E
+ \frac{11}{6} \right) f(\omega') \,. 
\end{align}
where the last equation also follows directly from the inverse Laplace transform of eq.~(\ref{eq:modelconv}).
Moreover, the inverse Laplace transform of the functions $\tilde K_i(s)$ yields
\begin{align}
    K_1(\omega) &= 1 - \frac{\alpha_s C_F}{4\pi} 
    \left( 2  \ln^2 \frac{\mu}{\omega} + 2 \ln \frac{\mu}{\omega}
    + 2 \,\textrm{Ei}(-\omega/{\hat\mu_F}) +\frac{\pi^2}{12} \right) 
    \,, \cr 
    \hat \mu_F \, K_2(\omega) &= 
    \, \frac{\alpha_s C_F}{4\pi} \left(
    2 \, \ln^2\frac{\mu_F}{\omega} - \frac{\pi^2}{3} \right) e^{-\omega/\hat \mu_F} 
    \,, \cr 
     K_3(\omega) &= \frac{\alpha_s C_F}{4\pi} 
    \left( 4 \, \delta(\omega) \ln \frac{\mu}{\mu_F} + 4 \, \frac{e^{-\omega/\hat \mu_F}-1}{\omega} -\frac92 \, \frac{ e^{-\omega/\hat \mu_F}}{\hat\mu_F} \right) 
    \,. 
\end{align}
Here, the exponential integral function is defined as the principal-value integral 
$$\textrm{Ei}(z) = - \int_{-z}^\infty dt \, \frac{e^{-t}}{t} \,.$$

\subsection{Asymptotic behaviour for large momenta}

\begin{figure}[t!b]
    \begin{center} 
    \fbox{\includegraphics[width=0.61\linewidth]{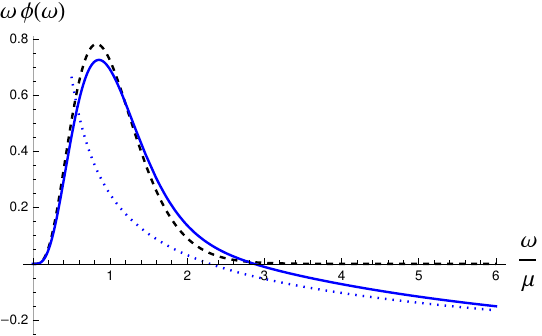}}
    \end{center}  
    \caption{Same as Fig.~\ref{fig:phi2tilde} 
    in momentum space: 
    The function $ \omega \, \phi(\omega) $ 
    as a function of  
    $\omega/\mu$.
    The black dashed curve shows the
    input model $f(\omega)$ in (\ref{eq:fmodel}). 
    The blue dotted line shows the 
    asymptotic behaviour in (\ref{eq:phi2tail}). 
    The solid blue line refers to the interpolation 
    following from  eq.~(\ref{eq:convolution}). 
    The input parameters are chosen as before, 
    $\alpha_s = 0.3$, $\mu e^{\gamma_E}= \mu_F e^{\gamma_E} = 2$~GeV
    and $\bar\Lambda = 0.6$~GeV, corresponding 
    to $\omega_0 \simeq 0.23$~GeV.}
    \label{fig:phi2}
\end{figure}

As the input function $f(\omega)$ vanishes exponentially for large arguments, the asymptotic behaviour of $\phi(\omega)$ 
can be obtained by expanding the kernel functions $K_i(\omega-\omega')$ for large $\omega \gg \omega'$. 
Since $K_2(\omega-\omega')$ vanishes exponentially, it does not contribute to the radiative tail.
We are then left with the asymptotic behaviour of the kernels $K_{1,3}$, with 
\begin{align}
     \int_0^\omega d\omega' \, 
    K_1(\omega-\omega') \, \frac{df(\omega')}{d\omega'} &=  
    \int_0^\infty d\omega' \, 
    \left( K_1(\omega) -\omega' \, K_1'(\omega) + \frac{\omega'{}^2}{2} \, K_1''(\omega) + \ldots \right) \frac{df(\omega')}{d\omega'} \,, 
\cr
\int_0^\omega d\omega' \, 
    K_3(\omega-\omega') \, f_{\rm aux}(\omega')
    &=  
    \int_0^\infty d\omega' \, 
    \left( K_3(\omega) -\omega' \, K_3'(\omega) + \ldots \right) f_{\rm aux}(\omega') \,.
\end{align}
\newpage
Here the moments of the input functions are calculated as 
\begin{align}
    \int_0^\infty d\omega' \left\{ 1, \omega', \frac{\omega'{}^2}{2} \right\} \frac{df(\omega')}{d\omega'} &=
\left\{ 
f(\omega)\bigg|_0^\infty\,, \ -\int_0^\infty d\omega'\, f(\omega')\,, \ -\int_0^\infty d\omega' \, \omega' \, f(\omega')  
\right\}
  \cr & = 
    \left\{ 0, - 1, - \frac{4\bar\Lambda_{\rm eff}}{3} + {\cal O}(\alpha_s) \right\} 
    \,, 
    \cr 
     \int_0^\infty d\omega' \left\{ 1, \omega' \right\} 
     f_{\rm aux}(\omega') &= 
     \left\{ 0 \,, \ 
     - \frac14 \, \int_0^\infty d\omega' \, \omega' \,  f(\omega')
     \right\} 
     \cr & =
     \left\{ 0, -\frac{\bar \Lambda_{\rm eff}}{3}\right\} \,,
\end{align}
where $\Lambda_{\rm eff}$ has been defined in (\ref{eq:fprimefix}).
Furthermore, for large values of $\omega$ we have
\begin{align}
K_1'(\omega) & \to \frac{\alpha_s C_F}{2\pi} \, \frac{1+2\ln \frac{\mu}{\omega}}{\omega} \,, 
\cr
K_1''(\omega) & \to - \frac{\alpha_s C_F}{2\pi} \, \frac{3+2\ln \frac{\mu}{\omega}}{\omega^2} \,,
\cr 
K_3'(\omega) & \to \frac{\alpha_s C_F}{\pi} \, \frac{1}{\omega^2} \,.
\end{align}
Putting the terms together 
determines the radiative tail of $\phi(\omega)$ at large values of $\omega$,
\begin{align}
    \phi(\omega) &= \frac{\alpha_s C_F}{2\pi} \, \frac{1}{\omega} 
    \left( 2  \ln \frac{\mu}{\omega} + 1 \right) 
    +  \frac{\alpha_s C_F}{2\pi} \, \frac{4\bar\Lambda_{\rm eff}}{3\omega^2} 
    \left( 2  \ln \frac{\mu}{\omega} + \frac72 \right) + \ldots  \,,
    \label{eq:phi2tail}
\end{align}
The result in eq.~(\ref{eq:phi2tail}) is analogous to the $B$-meson case
discussed in \cite{Lee:2005gza},
with only a different constant in the last bracket.

In Fig.~\ref{fig:phi2} we compare again the shape of the LCDA in momentum space as obtained from the treel-level input model (dashed line), the asymptotic 1-loop behaviour (blue dotted line) and the interpolation between the two (blue solid line). As one can observe, the effect of the radiative tail is two-fold. On the one hand, the LCDA crosses zero at large values of $\omega \sim 3\mu$. On the other hand, the peak of the distribution is slightly lowered. 
Again, these qualitative features have also been observed in the case of the $B$-meson LCDA in \cite{Lee:2005gza}.  
As already mentioned before, our approach to implement the radiative tail has the advantage that it does not change the analytic properties of the LCDA, and therefore our result for $\phi(\omega)$ yields a smooth function of $\omega$ (in contrast to the model that has been constructed for the $B$-meson LCDA in \cite{Lee:2005gza}).

\section{Conclusion}
\label{sec:conclusion}

In this paper, we have calculated the so-called ''radiative tail'' of the three-particle light-cone distribution amplitudes (LCDAs) of the $\Lambda_b$ baryon in
heavy-quark effective theory (HQET) from the short-distance expansion of the defining light-ray operators. 
Our one-loop results for the various matching coefficients of the dimension-9/2 and dimension-11/2 operators take a similar form as 
for the analogous case of $B$-meson LCDAs.
The corresponding short-distance behaviour of the $\Lambda_b$ LCDAs follows from the hadronic matrix elements which are expressed in terms of two ''decay constants'' and the HQET mass parameter $\bar \Lambda=M_{\Lambda_b}-m_b$.
We found it convenient to illustrate the radiative tail of the LCDAs
by introducing a systematic extrapolation to large distances, 
starting from a ''tree-level'' model that is used as input for a 
fixed-order integration of the renormalization-group equation 
with a sophisticated choice of integration constants.
For a given input model, this extrapolation takes a factorized form 
in position space which thus translates into a convolution in momentum space.
Not surprisingly, the qualitative behaviour of the $\Lambda_b$ LCDAs is found to be similar 
as for $B$-mesons.
Our results can be used to construct one-loop improved models for the $\Lambda_b$ LCDAs which enter 
theoretical predictions in QCD factorization or in the QCD sum-rule approach. 
This may further be used as the basis for phenomenological analyses of experimental results on 
exclusive $\Lambda_b$ decays in the future.
We also anticipate further applications of our idea to incorporate the fixed-order results for the radiative tail, 
e.g.\ generalizing our approach for generic parametrizations of $B$-meson LCDAs, see e.g.\ \cite{Feldmann:2022uok},
or for getting an improved description of the $B$-meson's three-particle LCDAs.

\subsubsection*{Acknowledgements}

We thank Björn Lange and Thomas Mannel for a critical reading of the manuscript and helpful comments and discussions.
This research  was supported by the Deutsche Forschungsgemeinschaft (DFG, German Research Foundation) under grant 396021762 - TRR 257.


\appendix

\section{\boldmath Renormalization-group equation for $\phi_2(\omega_1,\omega_2)$}

In momentum space, the renormalization-group equation for 
the $\Lambda_b$ LCDA $\phi_2(\omega_1,\omega_2)$ takes the 
form \cite{Ball:2008fw}
\begin{align}
 \frac{d\phi_2(\omega_1,\omega_2,\mu)}{d\ln\mu} &=
 - \left[\Gamma_{c}(\alpha_s) \,
  \ln \frac{\mu}{\sqrt{\omega_1\omega_2}} +\gamma_+(\alpha_s) \right] \phi_2(\omega_1,\omega_2,\mu)
 \cr
 & \qquad
- \frac{\omega_1}{2} \, \int_0^\infty d\eta_1 \, \Gamma_{+}(\omega_1,\eta_1,\alpha_s) \, \phi_2(\eta_1,\omega_2,\mu)
\cr & \qquad
- \frac{\omega_2}{2} \, \int_0^\infty d\eta_2 \, \Gamma_{+}(\omega_2,\eta_2,\alpha_s) \, \phi_2(\omega_1,\eta_2,\mu)
 \cr & \qquad
   + \frac{\alpha_s C_F}{2\pi} \, \int_0^1 dv \, V^{\rm ERBL}(u,v) \, \phi_2(v\omega,\bar v\omega,\mu) \,.
\label{baryon-one-loop}
\end{align}
Here the first three lines denote the Lange-Neubert evolution
\cite{Lange:2003ff},
where the leading terms in the various contributions to the anomalous dimensions in units of $\alpha_s C_F/4\pi$ are given by
\begin{align}
& \Gamma_{c}^{(1)} =4 \,, \qquad
\gamma_+^{(1)} = -2 \,, \qquad
\Gamma^{(1)}_{+}(\omega,\eta) = - \Gamma_{c}^{(1)} \left[ \frac{\theta(\eta-\omega)}{\eta(\eta-\omega)}
+ \frac{\theta(\omega-\eta)}{\omega(\omega-\eta)} \right]_+ \,,
\label{one-loop}
\end{align}
with the usual definition of the plus distribution.
In the last line $V^{\rm ERBL}$ is the Efremov-Radyushkin-Brodsky-Lepage kernel \cite{Efremov:1979qk,Lepage:1979zb}, arising from the gluon exchange between the two light quarks.

In the main text, we are considering the position-space LCDA at equal distances which corresponds to the integral over the momentum fraction $u=\omega_1/(\omega_1+\omega_2)$, for which the ERBL term in the above equation drops out. The position-space form of the Lange-Neubert terms at 1-loop has been discussed in \cite{Braun:2003wx,Kawamura:2010tj} in the context of discussing the RGE for 
the two-particle $B$-meson LCDA. Translated to the baryon case,
identifying 
\begin{align}
    \int_{-\infty}^\infty \frac{d\tau}{2\pi} \, e^{i\omega \tau} \, L \, \tilde \phi(i\tau) &= \ln \frac{\mu}{\omega} \, \phi(\omega) 
    - \int_0^\infty d\omega' \left[\frac{\theta(\omega-\omega')}{\omega-\omega'}\right]_+ \phi(\omega')
    \cr 
    &
    = 
    \int_0^\omega d\omega' \, \ln \frac{\mu}{\omega-\omega'} \,  \phi'(\omega')
    \,, 
\end{align}
and 
\begin{align}
     &\int_{-\infty}^\infty \frac{d\tau}{2\pi} \, e^{i\omega \tau} \,
     \int_0^1 dz \left[\frac{z}{1-z}\right]_+ \tilde \phi(i z \tau)
     = \phi(\omega) + 
     \omega \, \int_0^\infty d\omega' \left[\frac{\theta(\omega'-\omega)}{\omega'(\omega'-\omega)}\right]_+ \phi(\omega') 
    \,, 
\end{align}
we obtain eq.~(\ref{eq:ourRGE}) in the main text, with $\gamma_F^{(1)}= \gamma_+^{(1)}+\Gamma_c^{(1)}$.


\bibliography{refs}


\end{document}